\documentclass[sigconf,screen]{acmart}
\acmConference[ASE 2024]{IEEE/ACM Automated Software Engineering (ASE) Conference}{27 October - 1 November, 2024}{California, United States}

\usepackage{amsmath}
\usepackage[ruled,vlined,linesnumbered,boxed,commentsnumbered]{algorithm2e}
\usepackage{multirow}
\usepackage{algorithmicx}
\usepackage{booktabs}
\usepackage{graphicx}
\usepackage{makecell}
\usepackage{subcaption}
\usepackage{threeparttable}

\newcommand{\unnumberedfootnote}[1]{%
  \begingroup
  \renewcommand{\thefootnote}{}
  \footnotetext{#1}
  \addtocounter{footnote}{-1}%
  \endgroup
}

\copyrightyear{2024}
\acmYear{2024}
\setcopyright{acmlicensed}\acmConference[ASE '24]{39th IEEE/ACM International Conference on Automated Software Engineering }{October 27-November 1, 2024}{Sacramento, CA, USA}
\acmBooktitle{39th IEEE/ACM International Conference on Automated Software Engineering (ASE '24), October 27-November 1, 2024, Sacramento, CA, USA}
\acmDOI{10.1145/3691620.3695523}
\acmISBN{979-8-4007-1248-7/24/10}

\AtBeginDocument{%
  }

\begin{document}

\title{DevMuT: Testing Deep Learning Framework via Developer Expertise-Based Mutation}

\author{Yanzhou Mu\textsuperscript{$\dag$}, Juan Zhai\textsuperscript{$\ddag$}, Chunrong Fang\textsuperscript{$\dag$}\textsuperscript{$\ast$}, Xiang Chen\textsuperscript{$\#$}, Zhixiang Cao\textsuperscript{$\#$}, Peiran Yang\textsuperscript{$\dag$}, Yinglong Zou\textsuperscript{$\dag$}, Tao Zheng\textsuperscript{$\dag$}, Zhenyu Chen\textsuperscript{$\dag$}}

\affiliation{%
  \institution{\textsuperscript{$\dag$}State Key Laboratory for Novel Software Technology, Nanjing University}
  \country{}
}
\email{{602022320006,peiranyang,652023320004}@smail.nju.edu.cn,{fangchunrong,zt,zychen}@nju.edu.cn}

\affiliation{
  \institution{\textsuperscript{$\ddag$}University of Massachusetts Amherst}
  \country{}
}
\email{juanzhai@umass.edu}

\affiliation{
  \institution{\textsuperscript{$\#$}School of Artificial Intelligence and Computer Science, School of Zhang Jian, Nantong University}
  \country{}
}
\email{xchencs@ntu.edu.cn, zxcao@stmail.ntu.edu.cn}

%

\renewcommand{\shortauthors}{Mu et al.}
\begin{abstract}
  \unnumberedfootnote{$\ast$ Corresponding author} Deep learning  (DL) frameworks are the fundamental infrastructure for various DL applications. Framework defects can profoundly cause disastrous accidents, thus requiring sufficient detection. 
  In previous studies, researchers adopt DL models as test inputs combined with mutation to generate more diverse models. Though these studies demonstrate promising results, most detected defects are considered trivial  (i.e., either treated as edge cases or ignored by the developers).
  To identify important bugs that matter to developers, we propose a novel DL framework testing method DevMuT, which generates models by adopting mutation operators and constraints derived from developer expertise.  
  DevMuT simulates developers' common operations in development and detects more diverse defects within more stages of the DL model lifecycle (e.g., model training and inference).
  We evaluate the performance of DevMuT on three widely used DL frameworks (i.e., PyTorch, JAX, and MindSpore) with 29 DL models from nine types of industry tasks. 
  The experiment results show that DevMuT outperforms state-of-the-art baselines: it can achieve at least 71.68\% improvement on average in the diversity of generated models and 28.20\% improvement on average in the legal rates of generated models.
  Moreover, DevMuT detects 117 defects, 63 of which are confirmed, 24 are fixed, and eight are of high value confirmed by developers. 
  Finally, DevMuT has been deployed in the MindSpore community since December 2023. These demonstrate the effectiveness of DevMuT in detecting defects that are close to the real scenes and are of concern to developers.\looseness=-1
\end{abstract}

\maketitle

\section{Introduction}
\label{sec:introduction}

Deep learning (DL) applications have achieved widespread success in various industry applications, including safety-critical fields such as autonomous driving~\cite{ChenSKX15}, medical diagnosis~\cite{obermeyer2016predicting}, and facial recognition~\cite{voulodimos2018deep}. 
The development, maintenance, and deployment of such applications rely on DL frameworks. 
As such, framework defects can cause severe damages, leading to safety accidents and economic losses~\cite{teslanews}. 
Therefore, it is imperative to test DL frameworks and detect defects for their quality assurance.
Some researchers~\cite{pham2019cradle,guo2020audee,gu2022muffin,liu2023generation,wang2020lemon,li2023comet} adopt DL models as the test inputs and combine them with different mutation operators, such as modifying the model structure and editing layer parameters, to generate more diverse mutants and improve the test effectiveness. 

\begin{figure}[b]
     \centering
     \footnotesize
      \includegraphics[width=.95\linewidth]{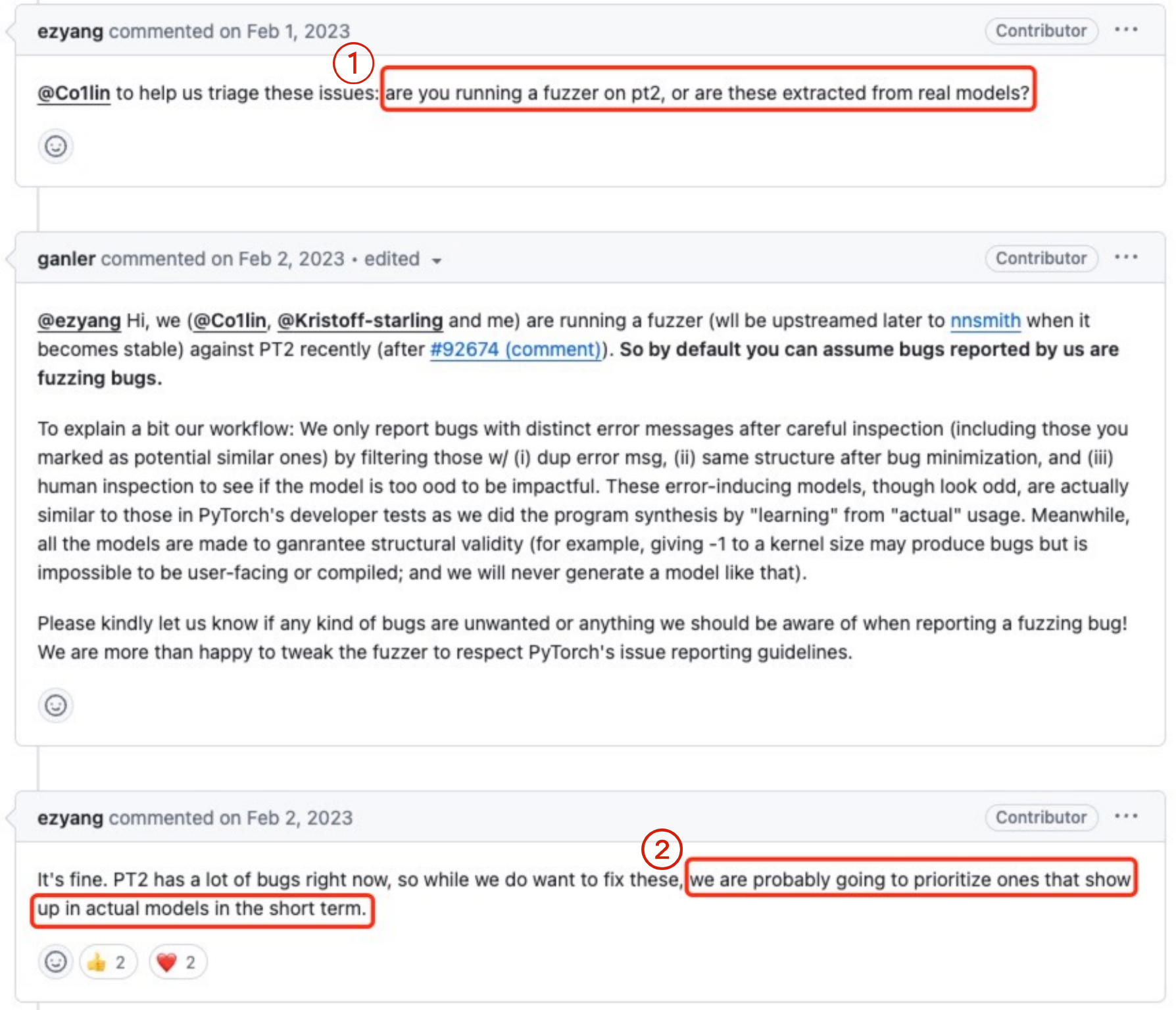}
      \vspace{-0.4cm}
     \caption{Developer Response on a Defect Report}\label{fig:demo}
     \vspace{-0.75cm}
\end{figure}


Although these methods have shown promising performance, they still struggle to detect the defects that developers are most concerned about.
Figure~\ref{fig:demo} shows an example of the response by PyTorch developers on a defect report~\cite{motivatingexample} submitted by a previous study~\cite{liu2023neuri}. 
It generates an edge model that is rare in the real world and detects an inconsistency defect about the group of PyTorch's ``Linear'' and ``Batchnorm2d'' operators. 
Developers' responses in \textcircled{1} and \textcircled{2} show that they prefer fixing defects in actual models over those generated by fuzzers. 
Even if the submitted defect is confirmed, developers usually ignore it since the model is rare in real scenes. 
In other words, the DL framework community mainly focuses on defects that profoundly threaten actual models applied in real scenes, such as unreasonable resource scheduling, abnormal performance, and substandard efficiency. 
This demonstrates the gap between developers and previously proposed methods. \looseness=-1

More specifically, previous work's focus is misaligned with the developers' community and they suffer from three main limitations from the perspective of DL framework developers.
First, \textbf{they detect impractical defects.}
Existing solutions often modify the structure, parameters, and neuron weights to generate more diverse DL models, neglecting whether these modifications occur in real development and whether the generated models reflect actual cases in the real world.
Though the generated models can expose defects, the triggered defects are rare in the real world and often automatically closed without any solutions after confirmation.
Second, \textbf{they overlook critical defects.}
They only focus on whether the generated models can execute normally and achieve consistent results across different frameworks in model inference while ignoring other subjects (e.g., resource usage) in other stages (e.g., training) of model lifecycles. 
It leads to missing the defects related to unreasonable resource scheduling, substandard efficiency, abnormal performance, etc, thus limiting the detection ability. 
These defects are more frequently exposed during development and often interfere with the users when using frameworks, which is also a concern for developers.
Finally, the last limitation is \textbf{the consumption of massive resources}.
They lack constraints to filter illegal models that may lead to crashes, produce abnormal outputs not caused by framework defects, and trigger abundant false positives. 
This wastes time analyzing low-value test inputs and requires extra manual efforts to identify false positives, thus reducing test efficiency. \looseness=-1

To bridge the gap, we propose a novel DL framework testing method DevMuT, focusing on effectively exploring valuable test input space and detecting diverse defects based on developer expertise. 
Specifically, we collect practical experience in developing DL models from our industry partners and design seven mutation operators and nine constraints to avoid detecting impractical defects and save efforts caused by illegal models. Particularly,
the mutation operators are designed to simulate the common operations of the developers on DL models, and the mutation constraints are adopted to guide how to mutate models and filter models that are meaningless in detecting defects.
To detect more diverse and critical defects, we consult our industry partners about the main subject (e.g., resource usage) and defect type they care about during the model construction, training, and inference to design corresponding test oracles and detect diverse defects in these stages. 
In particular, we analyze common resources (such as memory, GPU, execution efficiency, and model performance) to examine differences across different frameworks. Besides, we detect whether there are crashes or abnormal outputs during execution.
Finally, to further improve the effectiveness, we utilize the double-Q learning algorithm~\cite{hasselt2010double}, a popular reinforcement learning algorithm to select mutation operators and seed models during the mutation process. \looseness=-1


To evaluate DevMuT, we collect 29 models as seeds from nine kinds of industrial tasks. 
We apply DevMuT on three popular DL frameworks (i.e., PyTorch~\cite{torch}, JAX~\cite{jax}, and MindSpore~\cite{mindspore}) with the above DL models as the test inputs.  
Compared with the state-of-the-art baseline methods~\cite{li2023comet,gu2022muffin}, DevMuT achieves an average increase of 71.68\% on the generated model diversity. 
Meanwhile, DevMuT can effectively generate legal models with an 81.17\% legal rate with an improvement of 28.20\% compared with baselines.
Moreover, DevMuT detects 117 defects, with three new kinds of defects not detected by baselines (i.e., six abnormal memory usage defects, six abnormal performance defects, and ten train crash defects). 
Among them, developers have confirmed 63 defects. 24 defects have been fixed, and eight defects are labeled with specific tags such as ``Main'', representing the defects that receive enough attention from developers. We also successfully contributed a pull request about fixing defects in MindSpore.

Our contributions can be summarized as follows: \looseness=-1

\begin{itemize}
    \item \textbf{Perspective.}
    We are the first to leverage developer expertise in designing mutation operators and constraints. Additionally, we extend the detection scope to include the construction and execution stages (including model training) of the model lifecycles rather than only the inference stage.
    
    \item \textbf{Approach.} We design a practical DL framework testing method DevMuT, which adopts the preset mutation operators and constraints combined with double-Q learning to effectively generate new models common in real scenes and detect more diverse defects.
        
    \item \textbf{Evaluation.} We collect a large-scale benchmark and conduct extensive experiments to evaluate the effectiveness of DevMuT. The results show that DevMuT detects 117 defects (63 confirmed) with three new types of 22 defects that baselines cannot detect. It also detects eight defects labeled with specific tags like ``Main'' to indicate higher priority for developers. \looseness=-1

    \item \textbf{Practicality.} DevMuT has been applied in the MindSpore community~\cite{devmuttest} for continuous quality assurance since December 2023. We also implement DevMuT and share our proposed method on GitHub~\cite{sharelink} for open science.
    

\end{itemize}

\section{Background and Related Work}
\label{sec:background}

 
\subsection{DL Model and DL Framework}
\label{sec:background2.1}

\begin{figure}[]
     \centering
      \includegraphics[width=0.95\linewidth]{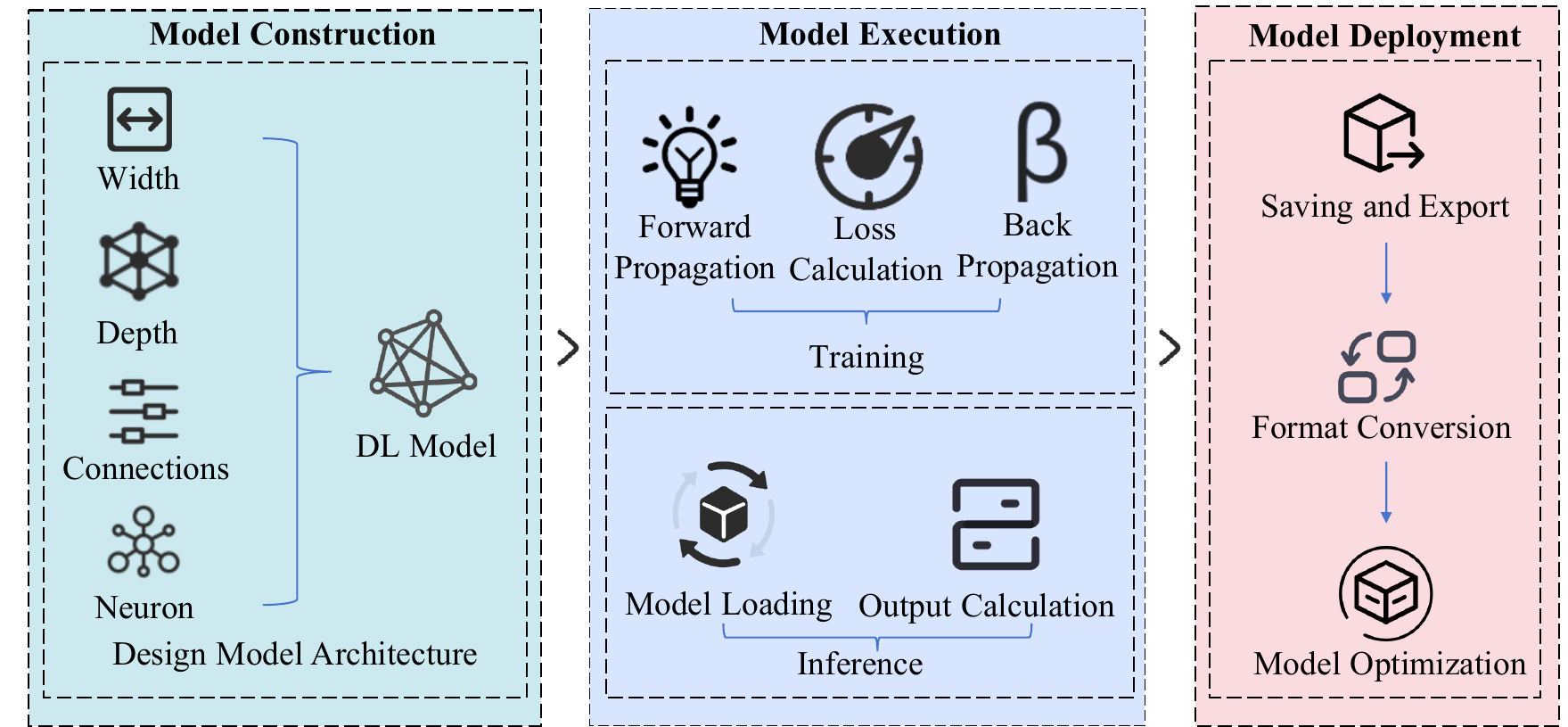}
      \vspace{-4mm}
     \caption{The Lifecycle of DL Models}
    \vspace{-8mm}
          \label{fig:dlstrcture}
\end{figure}

DL model comprises multiple middle layers consisting of one or more DL operators, such as convolution, pooling, and activation. These layers contain many neurons, and the connections between layers depend on the different weights of the neurons. These weights are updated with the external input data to fit specific scene tasks like image classification. 
The whole lifecycle of DL models includes the construction, execution, and deployment as shown in Figure~\ref{fig:dlstrcture}. 
These stages require fundamental support from DL frameworks to ensure performance and correctness. Framework defects exposed in the model lifecycle are more diverse and highly relevant to actual scenes. Therefore, we introduce the model lifecycle and the correlations with DL frameworks. 
Developing DL models involves designing the architecture based on tasks, including the depth, width, neurons, and connections of different middle layers. DL frameworks encapsulate common-used functions as DL operators and hide the implementation complexity~\cite{fayad1997object} for implementing the model architecture~\cite{faught1986applications}. 
Model execution involves training and inference processes.
The training process includes forward propagation, loss calculation, and back-propagation:
(1) forward propagation: the input data sequentially passes through the middle layers and calculates with the neuron weights; 
(2) loss calculation: it calculates the difference between the model outputs and the real labels through the loss function to evaluate the accuracy; 
(3) back-propagation: the model updates the weights from the last output layer and reverses to the first input layer by adjusting the gradients of each weight and optimizing to reduce the loss.
The three steps are repeated to reduce losses and improve model performance until it stabilizes.
After training, the model performs inference on new data by executing forward propagation and conducting predictions.
DL frameworks also implement the functionalities for these computing tasks to improve efficiency and simplify execution scripts.
Moreover, DL frameworks efficiently schedule computing resources for smooth execution and optimize global resource scheduling to ensure stable and efficient execution under parallelism requirements.
Deployment involves saving and exporting the trained DL models, converting them into specific formats, ensuring compatibility with the target environment, and optimizing for inference speed, resource use, and stable execution. DL frameworks efficiently support cross-platform compatibility for different environments and enhance inference speed and resource utilization through optimization techniques. \looseness=-1


\vspace{-2mm}
\subsection{Testing DL Frameworks via DL Models}

Existing works that adopt DL models as test inputs for DL frameworks can be divided into two types: (1) one kind of method utilizes mutation operators to mutate existing DL models in public repositories~\cite{pham2019cradle,guo2020audee,wang2020lemon,li2023comet}; (2) another kind of methods generate models based on predefined templates and constraints~\cite{gu2022muffin,liu2023neuri,liu2023generation,luo2021graphfuzz,wang2022eagle}. \looseness=-1

\noindent\textbf{Mutation-Based Testing Methods.} 
Pham et al.~\cite{pham2019cradle} proposed CRADLE, the first model-level testing method that utilizes differential testing by analyzing model execution on different backends of Keras~\cite{keras} (e.g., TensorFlow~\cite{tensorflow}, CNTK~\cite{cntk}). 
Wang et al.~\cite{wang2020lemon} proposed LEMON and adopted mutation operators in DeepMutation~\cite{ma2018deepmutation} and DeepMutation++~\cite{hu2019deepmutation++} combined with the MCMC strategy, which has been widely used in many fields~\cite{chen2015guided,Chen2016Coverage2} to guide the mutation process. It utilizes differential testing on the generated mutants to detect defects. 
Guo et al.~\cite{guo2020audee} proposed Audee, which combined genetic evolution strategy~\cite{vidnerova2016evolutionary} with input-mutation to generate mutants that can trigger higher inconsistency and outlier output. 
Li et al.~\cite{li2023comet} proposed COMET and added new types of mutation operators, such as parameter-mutation and outlier mutation, compared with LEMON. Its execution process is similar to LEMON.

\noindent\textbf{Generation-Based Testing Methods.} 
Gu et al.~\cite{gu2022muffin} proposed Muffin, which generates new models based on the structure of the directed acyclic graph. It focuses on detecting inconsistency defects in three model execution processes: forward propagation, loss calculation, and back-propagation calculation. 
Luo et al.~\cite{luo2021graphfuzz} proposed Graphfuzz to test DL inference engines based on graph theory. It combines six mutation operators with the MCTS algorithm~\cite{Coulom2006EfficientSA} to explore the generation of new model structures. 
Liu et al.~\cite{liu2023generation} proposed Gandalf, which generates different models using context-free syntax combined with the DQN algorithm~\cite{mnih2015human}. It introduces 15 metamorphic relationships to enhance the testing effectiveness of different DL frameworks. 
Liu et al.~\cite{liu2023neuri} proposed Neuri, which generates models through inductive rules. It first collects the call relationships of framework interfaces from external resources and analyzes legal interface groups to generate test inputs that can explore the deeper execution behavior of DL frameworks.

\noindent\textbf{Comparison with Previous Studies.} As the mutation-based method, DevMuT shows the following novelties.
(1) The mutation operators adopted in existing methods are directly from the DL model testing work~\cite{ma2018deepmutation,hu2019deepmutation++,humbatova2021deepcrime}. In contrast, the mutation operators in DevMuT are abstracted from the common operations of developers on DL models.
(2) Existing methods fail to filter out illegal models, while DevMuT goes a step further by filtering them based on developer tactics.
(3) Existing work focuses on detecting the inconsistency defects exposed in model inference while DevMuT detects other kinds of defects from the lifecycle of DL models closer to real scenes.\looseness=-1

\section{Mutation Design Based on Developer Expertise} 
\label{sec:mutationdesign}

To detect defects that are more relevant to real scenes, we interview developers to gather their expertise in developing DL models and framework defects. 
Based on the results, we aim to design mutation operators to simulate their common operations in development and design mutation constraints abstracted from development tactics to guide the mutation and filter illegal models.
Next, we introduce the interview process, present the interview result analysis, and conclude the findings. Finally, we further introduce the design of the mutation operators and constraints based on the findings.


\vspace{-2mm}
\subsection{Interview Process}

We design an open guideline to clarify our background and interview process with different types of questions, e.g., short answer, choice, and open-ended questions. Details can be found on our website~\cite{sharelink}. Here are five aspects of interview questions. \looseness=-1

\noindent\textbf{\underline{Part I: }} We ask them five short questions about their basic profiles including jobs, duties in DL framework development, and working experience in years.
\looseness=-1

\noindent\textbf{\underline{Part II: }} We ask them three short questions and one multi-choice question about the subjects (e.g., model scripts) they often process and their specific operations on these subjects.

\noindent\textbf{\underline{Part III: }} We ask them five short answer questions about their experience in identifying and dealing with those weird models that cannot fit into real scenes.

\noindent\textbf{\underline{Part IV:}} We ask them six short answer questions about the defect they care about with a specific introduction about the damage to software, symptoms, root causes, trigger frequency, etc. We also asked how they evaluate the importance of defects.

\noindent\textbf{\underline{Part V: }} We ask them six open-ended questions about their opinions on existing frameworks and expectations for optimizations.\looseness=-1


We invite six professionals from leading Internet companies as interviewees who are our industry partners, specialize in DL framework quality assurance, and are responsible for developing, testing, and detecting defects in new releases. 
With over four years of experience in defect detection, localization, and fixing, the interview quality can be guaranteed. We begin by introducing the guidelines and providing background information on our research. We inform the participants that the interview will be recorded, obtain their approval, and ensure confidentiality. Each interview is conducted online and lasts approximately 30 minutes. After the interview, we verify that no sensitive information is disclosed and send our gratitude for their participation.

\noindent\textbf{Results Analysis.}
After all interviews, one author transcribes records and extracts the development operations and defect types. The other three authors verify the results and provide optimization suggestions. All the authors then discuss different opinions until reaching a consensus. We conclude with three main findings from the interview results. \looseness=-1


\noindent\textbf{Findings about the Common Operations in Development.} 
The operations refer to the developers' modifications on various model scripts. 
We find two types of scripts that developers frequently modify: 
(1) structure scripts that design the architecture of models and 
(2) execution scripts that control the training and inference. 
For the structure scripts, we find that developers often introduce new structures to replace existing structures, add them to original models, delete existing structures, and modify the parameters to improve model performance. 
For the execution scripts, we find that developers often modify data, loss function, and optimizer to enhance the generalization ability and improve convergence speed. Specifically, they often conduct data augmentation, change the loss function and optimizer type, and modify their parameters.
These findings motivate us to identify the mutation objects and operators, i.e., we adopt the structure script and execution script as the mutation objects while the operations are implemented as mutation operators. \looseness=-1


\noindent\textbf{Findings about the Tactics in Development.}
The tactics are (1) the notes when modifying scripts and (2) the experience of identifying illegal models.
For the first kind of tactic, developers provide six guidelines, covering requirements for adding or replacing new structures, shape dimensions, parameter setting ranges, and the position for adding or deleting middle layers. 
For the second kind of tactic, developers suggest three tactics, ranging from time-consuming to output accuracy and gradient value, to identify illegal models.
These findings motivate us to design the constraints to guide the common operations and identify those weird models after modifications.\looseness=-1

\noindent\textbf{Findings about the Defects Developers Care About.}
We conclude the subjects that developers often monitor to detect defects and the relevant defect types.
Specifically, developers often focus on resource usage (e.g., CPU, memory, and GPU), model performance such as training loss and inference evaluation, execution efficiency, and output accuracy in model execution. The defects related to these subjects (e.g., substandard performance and crashes) achieve higher fix priority of developers since they believe these defects frequently appear in real-world scenarios and significantly impact users.
The findings guide us to design test oracles to detect defects and are introduced in Section~\ref{sec:bugdetection}.

We present these findings to our interviewees for confirmation and obtain their approval.
Moreover, they give corresponding suggestions about defect detection. All the details of the interviews can be found on our website~\cite{sharelink}.
In summary, we design seven mutation operators to simulate the common operations of developers on the structure and execution scripts based on the above results. Meanwhile, we also design six constraints to guide mutation and three constraints to filter illegal mutants abstracted from the tactics. 
Besides, we focus on four types of defects as recommended by developers with ten test oracles for detection.
The details of mutation operators and constraints are in the following part of this section while the defects and the test oracles are in Section~\ref{sec:bugdetection}.  \looseness=-1




\vspace{-4mm}
\subsection{Mutation Operator Design} 
We design the mutation operators based on interview insights from three aspects. 
\textbf{Modified Objects.} Developers frequently modify the model structure and execution script, focusing on middle layers, layer parameters, and loss/optimizer functions to enhance performance. This leads us to target these objects for mutation.
\textbf{Operator Type.} Developers often add, delete, or replace middle layers, and expand or reduce data outputs or layer parameters. We implement mutation operators to simulate these actions.
\textbf{Mutation Guidelines.} Developers adhere to guidelines, such as not placing one Conv2D layer immediately after another, to maintain model performance. We incorporate these constraints to ensure the validity of mutations. The details about the mutation operators are as follows.
We adopt (1) the model structure script and (2) the execution script as mutation subjects and propose seven mutation operators (MOs). \looseness=-1

As the first kind of mutation subject, the model structure can be divided into three types:
(1) backbones for extracting features from input data such as the VGG~\cite{simonyan2014very}, ResNet~\cite{he2016deep}, Inception~\cite{sun2014ranking}; 
(2) cascade operators consisting of multiple DL operators for executing specific tasks (e.g., the RPN~\cite{ren2015faster} structure for identifying target area in object detection); 
(3) basic operators that cannot
be further subdivided, such as Conv2d, etc. 
The five kinds of common operations on the above three types of structures are as follows and each operation represents one type of mutation operator.

\noindent\textbf{MO1: Existing Structure Replacement.} 
This mutation operator refers to inducing external model structures (such as new backbone, cascade operators, and basic operators) to replace the existing structures in the seed model. 

\noindent\textbf{MO2: Shape and Dimension Change.} 
This mutation operator refers to adjusting the input/output size or dimension of the basic operators or the first and last of cascade operators. 
Developers enlarge dimensions or shapes to enhance learning on complex data, reduce them to lower computing costs, and improve generalization ability. \looseness=-1

\noindent\textbf{MO3: Series Connection Addition/Deletion.} 
This mutation operator refers to changing the depth (i.e., the total number of middle layers) of DL models by adding or deleting existing cascade or basic operators that are ordered in series connection. Developers often insert new structures to enlarge the model depth, learn the data feature deeper, and delete replicate structures for higher efficiency. \looseness=-1

\noindent\textbf{MO4: Parallel Connection Addition/Deletion.} 
This mutation operator refers to changing the width (i.e., the total number of neurons in each layer) of DL models by adding or deleting branches consisting of cascade operators or basic operators. 
Developers add new branches to capture multi-dimensional data features, enhance model robustness, and enable parallel training. Besides, They also prune branches to avoid overfitting and simplify computation. 

\noindent\textbf{MO5: Parameter Value Change.} 
This mutation operator refers to modifying the parameter values of the basic operators.
Developers modify layer parameters to adjust the model's width and depth, avoid overfitting, and adapt to new tasks, as changes in these parameters influence performance.

As the second kind of mutation subject (i.e., the execution script), the main objects include the loss function, optimizer, training/test set, and parameters to decide training or inference execution. We design the following two mutation operators on loss function and optimizer as recommended by developers.

\noindent\textbf{MO6: Loss Function Change.} This mutation operator involves the loss function's type and parameter value change, as developers often do it to adapt to different kinds of new tasks.

\noindent\textbf{MO7: Optimizer Change.} This mutation operator involves the optimizer's type and parameter value change, as developers often do to promote the training process. 

Finally, we count their proportion mentioned by interviewees: MO1, MO2, MO3, MO4, MO5, MO6 and MO7 are mentioned by 100\%, 50\%, 83.3\%, 83.3\%, 66.7\%, 50\% and 50\% interviewees, respectively.

\subsection{Mutation Constraint Design} 
\label{sec:mcdesign}

We collect the tactics for modifying models and estimating whether the models are legal from the interview results. Specifically, we design two types of constraints (MCs): (1) to guide mutation on seed models and (2) to filter illegal models.
Such MCs are adopted to reduce false positives and improve the efficiency of DevMuT. We introduce these MCs and show the details on our website~\cite{sharelink}.

\noindent\textbf{MC1: Structure Functionality Consistency Constraints.} 
The new structures should have the same functionality to keep the consistency of the target model (e.g., replace a normal convolution with a depthwise separable convolution) since replacing new structures of different functions (e.g., replace a normalization layer with an activation layer) may disrupt the stability, increasing execution costs and reducing the interpretability. We collect five kinds of backbones (e.g., ResNet) to extract image data features, seven kinds of cascade operators (e.g., residual block) to enhance the feature learning, and five kinds of basic operators (e.g., avgpool and maxpool operators) with different functionalities for candidate replacement. 


\noindent\textbf{MC2: Data Calculation Scale Constraints.}
We set the change rate of shape or dimension from one-fourth of the original shape or dimension to four times as recommended by developers to avoid excessive computing costs.

\noindent\textbf{MC3: Series Structure Addition/Deletion Constraint.}
We only add new cascade operators with the same task type and avoid adding the same basic operators at the same position (e.g., add one ``Conv2d'' after one ``Conv2d'').
Besides, we only select one of the multiple similar structures to delete (e.g., delete the one entire ``Conv2d-Batchnorm2d-Relu'' structure in VGG) for the functionality consistency between the mutation model and the origin. According to the developer's recommendation, the range for adding or removing series structures for common image classification and object detection models should not exceed 20\% of the original model depth. \looseness=-1

\noindent\textbf{MC4: Parallel Structure Addition/Deletion Constraint.}
Similarly to MC3, the new branch structure should share the same functionality as the original structure.
Besides, the same position cannot be added to more than two branches to reduce time costs.

\noindent\textbf{MC5: Parameter Value Range Constraints.}
The change value range is decided based on their effects on model performance. We select 12 kinds of parameters from four basic operators as the mutation range followed by the recommendation of developers. For example, for the parameter ``kernel\_size'' of the ``Conv2d'' operator, we follow the developer's recommendation and adopt three different sizes: ``3*3'', ``5*5'' and ``7*7'' to adapt to different data features.
The other value change range can be found on our website~\cite{sharelink}.

\noindent\textbf{MC6: Loss/Optimizer Modification Constraints.}
The types of loss functions and optimizers with their parameters are decided based on the model tasks. We collect four kinds of optimizers and four kinds of loss functions as the candidates.  

Besides, we propose three filter constraints to filter weird models: \looseness=-1

\noindent\textbf{MC7: Execution Time Constraints.}
Developers suggest that the acceptable range for the increased training time of the generated model is less than two to three times and the increased inference time is less than one to two times compared with the original. We take the inference time as the standard and set the threshold three times to eliminate external interference. 

\noindent\textbf{MC8: Output Accuracy Constraints.} The output value exceeding the accuracy range (e.g., the accuracy range of MindSpore is under 1e38 at ``float32'') is meaningless and produces outliers like ``NAN'' ``inf''. This may induce more false positives and reduce test effectiveness. We adopt the accuracy range of ``float16'', ``float32'' and ``float64'' as the constraints to filter those models with invalid output. \looseness=-1

\noindent\textbf{MC9: Gradient value range Constraints.} 
If the output gradient of the generated model contains outliers (exceeds 1e2 or is lower than 1e-3), it is filtered.
Such models do not change performance since the early training stage and are worthless for analyzing their execution.\looseness=-1

\section{Approach}
\label{sec:method}

\vspace{-1mm}
\subsection{Overview}

\begin{figure*}[]
     \centering
    \includegraphics[width=0.95\linewidth]{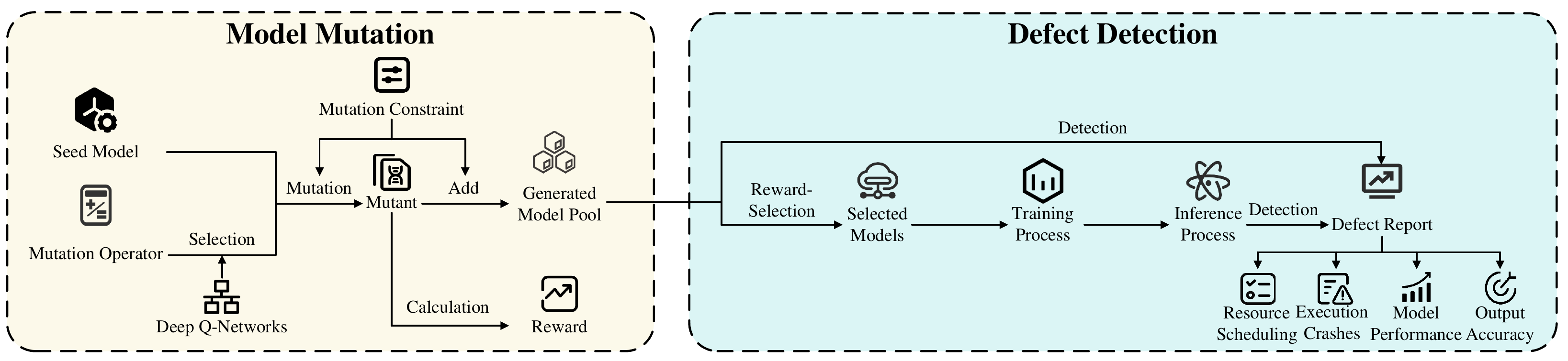}
      \vspace{-0.4cm}
     \caption{Workflow of DevMuT}   
     \vspace{-0.4cm}
     \label{fig:workflow}
\end{figure*}

Figure~\ref{fig:workflow} presents the workflow of DevMuT. 
It consists of two parts: \textit{Model Mutation} that generates mutants, and \textit{Defect Detection} that performs defect detection.
In \textit{Model Mutation}, DevMuT adopts two deep Q-networks to select mutation operators and seed models. 
DevMuT mutates the seed model under the guideline constraints (MCs 1-6 in Section~\ref{sec:mcdesign}) and updates the Q-networks based on the reward calculated from the evaluation of the current mutation. 
It filters out mutants that violate constraints (MCs 7-9 in Section~\ref{sec:mcdesign}) and adds legitimate ones into the generated model pool.
The process iterates until mutation finishes (Section~\ref{sec:modelmutation}).
In \textit{Defect Detection}, DevMuT first detects and records the defects exposed in mutants.
Based on rewards, DevMuT selects part of the models and further executes them to detect defects in resource scheduling, execution crashes, model performance, and output accuracy (Section~\ref{sec:bugdetection}). 

\begin{algorithm}[t]
	\scriptsize
	\SetAlgoVlined
	\SetKwInOut{Input}{\textbf{Input}}\SetKwInOut{Output}{\textbf{Output}}
	\Indm
	\Indp
	\Input{$M$: the original DL model; \
		$S_{pool}$: the mutation operator pool;\
		$N$: the number of mutation execution rounds;\
        $k$: the selected number of models;\
		}
	\Output{$DefectSet$: the set of all the detected defects;}
	\BlankLine
 
	$n \gets 0$, d $\gets M$, $DefectSet$ $\gets \emptyset$, $D$ $\gets \{M\}$, init($Q_1$, $Q_2$);\\
    \While{n $\leq$ N}{
    
        $p$ $\gets$ random.uniform(0, 1);\\
        \eIf{$p$ $\leq \epsilon$}
        {
            s $\gets random.choice(S_{pool})$;\\
        }
        {
                s $\gets S_{pool}[\arg \max (\frac{Q_1(d, *) + Q_2(d, *)}{2})]$;   
        }
        
        $d'$ $\gets$ s($d$); \\ 
        d.select\_num++, d.snums[indexof(s)]++;\\
        
        \eIf{judge($d'$) == 1}{
		  	r $\gets$ -1, $d'$.r $\gets$ r;\\
                $d'$ $\gets UCBSort(D)$;\\
		}
        {
            r $\gets$ $distance(d',d)$, $d'$.r $\gets$ r;\\
            \If{ $d'$ not in $D$}
            {
                $D$.add($d'$.copy());\\
            }
            
        }
        p $\gets$ random.uniform(0, 1);\\
        \eIf{p $\leq$ 0.5}
        {
           s' $\gets S_{pool} \arg \max(Q_1(d', *))$ 
           $Q_1.update(Q_1(d, s) + \alpha [r + \gamma Q_2(d', s') - Q_1(d, s)] )$
        }
        {
            s' $\gets S_{pool} \arg \max(Q_2(d', *))$ 
            $Q_2.update(Q_2(d, s) + \alpha [r + \gamma Q_1(d', s') - Q_2(d, s)] )$
        }
        d $\gets$ d'.copy(), n++;\\
        }

        \ForEach{$d$ in D}{
            ifbug $\gets$ $detect\_{bug}$($d$, $DefectSet$, $"Mutation"$);\\
            \If{ifbug}
            {
                $D$.remove($d$);\\
            }
        }
        
        $D$ $\gets Sort(D)[:k]$;\\
       
       \ForEach{$d$ in $D$}{
            $detect\_{bug}$($d$, $DefectSet$, $"Execution"$);\\
        }
        
        \Return $DefectSet$;\\

	\caption{The DevMuT Algorithm }
    
	\label{alg:DevMutTest}
\end{algorithm}

\subsection{Model Mutation}
\label{sec:modelmutation}

This section introduces how DevMuT mutates models as shown in Algorithm~\ref{alg:DevMutTest} (Lines $1-22$) while the core is to select mutation operators and seed models. DL models contain complex structures and numerous parameters and are more complicated after mutation, further increasing the exploration difficulty of the test input space.
To overcome this challenge, DevMuT adopts double-Q learning~\cite{hasselt2010double}, a reinforcement learning strategy that utilizes two deep Q-networks for selection. It can efficiently identify valuable parts of the test input space and continuously explore it with two Q-networks. \looseness=-1


The problem in our scene is modeled using reinforcement learning (RL) to explore the test input space. Key elements of RL are ``state'' which represents the current situation of the environment, ``action'' which represents the transformations between two states, and ``reward'' which represents the feedback received from the environment. 
In our scenario, ``state'' is the generated model, while ``action'' is the mutation operator. 
Since we aim to generate more diverse models and different from the original model, we adopt the output distance between the generated and original models as the ``reward''. The workflow based on the above definition is shown in the following parts.\looseness=-1

DevMuT first initializes the parameters  (Line $1$):
(1) the selected seed model $d$ to mutate;
(2) two networks $Q_1$ and $Q_2$ for selecting mutation operators;
(3) the set $DefectSet$ that stores detected defects;
(4) the mutation counter $n$;
(5) the set $D$ stores the generated models. 
Then it executes $N$ rounds of mutation (Lines $2$-$22$).
To effectively select mutation operators, DevMuT utilizes the $\epsilon$-greedy policy~\cite{mnih2015human} strategy: it generates a probability $p$ from the symmetric distribution (0,1) (Line $3$) and compares it with the preset threshold $\epsilon$ (Line $4$). If $p$ is smaller than $\epsilon$, DevMuT randomly selects one mutation operator (Line $5$). Otherwise, it calculates the output average of $Q_1$ and $Q_2$ for selecting the mutation operator $s$ (Line 6-7). The probability $p$ determines the likelihood of selecting a mutation operator strategy. Operators are more likely selected based on the output of two Q-networks with larger $p$ values, as this double Q-network strategy explores a more diverse test input space.
Then DevMuT mutates $d$ with $s$ to generate the model $d'$ (Line $8$). 
If the current mutation fails (e.g., generate crash models)
or generates those illegal models that violate three constraints (Line $10$), DevMuT sets the reward $r$ to -1 for punishing the selection of $s$ based on $d$ (Line $11$) and selects a new seed model with the maximum value of $UCB$ (Line $12$) and assign it to $d'$.
$UCB$~\cite{auer2002finite} is often adopted to avoid trapping into the local optimum and give more chances to explore new space. The definition of $UCB$ is shown as follows:\looseness=-1
\vspace{-2mm}
\begin{equation}
    UCB(d) = d.r + c * \sqrt{\frac{\ln{(d.select\_num)}}{d.snums[indexof(s)]}}
    \label{equ:ucb} 
\end{equation}
\vspace{-4mm}

In the above formula, $d.r$ represents the current reward of $d$, $d.select\_num$ represents the selected counts of the seed model, 
$d.snums$ represents the selected counts of $s$ in $d$, 
$c$ is the weight to balance the left and right parts, respectively. $d.select\_num$ and $d.snums$ are updated after the mutation (Line $9$).

If the generated model $d'$ does not violate the constraints (Line $13$), DevMuT calculates reward $r$ between the current seed model $d$ and $d'$ (Line $14$) and update the model pool $D$ (Lines 15-16). 
The definition of the reward is as follows: Given two models $D_1, D_2$ with $n$ layers $ f= \langle L_0, L_1, \cdots, L_n \rangle$ and an input tensor $x$, the output of the $i$-th layer is recorded as $ f_{L_i} (x)$. We adopt Chebyshev distance~\cite{cantrell2000modern} for calculating the distance between $D_1$ and $D_2$ on layer $i$:
\vspace{-2mm}

\begin{equation}
    r^{D_1,D_2}_{f_{L_i}}(x) = mean (|D_1{_{f_{L_i}}} (x) - D_2{_{f_{L_i}}} (x) |)
    \label{equ:layerdis} 
\end{equation}

We adopt the final layer as the basis for calculating reward. Then, DevMuT randomly updates one of the two Q-networks (Lines 17-21). 
We follow the original update process of double Q-learning, i.e., combine the currently expected reward $Q_1(d, s)$ or $Q_2(d, s)$ (left part of the Lines $19$ and $21$) with the future expected reward $Q_1(d', s')$ or $Q_2(d', s')$ (right part of the Lines $19$ and $21$). The parameters $\alpha$ and $\gamma$ are adopted to control the update speed and measure the contribution of the future reward, respectively. Finally, DevMuT utilizes $d'$ as the seed model of the next round (Line $22$).

\subsection{Defect Detection}
\label{sec:bugdetection}

This section introduces DevMuT for defect detection and presents the test oracles. As shown in Algorithm~\ref{alg:DevMutTest} (Lines 23-29),
DevMuT further detects defects within mutation by analyzing the models in set $D$ that stores generated legal models (Line $24$) and removes the defective models from $D$ (Lines 25-26). 
DevMuT then sorts $D$ based on the reward value and selects the top $k$ models (Line $27$) for further analysis. 
It further trains the selected models and executes inference to detect defects (Lines 28-29). Finally, DevMuT returns $DefectSet$ as the output (Line $30$).
Since these subjects have different value ranges on different models due to the input data, scene task types, etc, we run all the models across different frameworks to collect preliminary results (e.g., memory usage size) about these subjects. Then, we further set the thresholds for determining the generated models expose the relevant defects. It also records crashes during execution. We use a table to show the thresholds of each model about these subjects, which can be found on our website~\cite{sharelink}. The test oracles for determining whether these subjects expose defects are as follows. 

\noindent\textbf{Performance Defect.}
The performance involves the training loss, execution efficiency, and inference evaluation. 
We design two test oracles about loss:
(1) the model outputs an abnormal loss value (e.g., NAN and inf) on one framework but not on another;
(2) when the loss value is normal, we adopt dynamic time warping (DTW) distance~\cite{muller2007dynamic} to measure the loss similarity of the models across different DL frameworks and report defect if the DTW distance exceeds the preset threshold. It is commonly used in speech recognition~\cite{xiong2023fundamentals} to measure the similarity between two time series.
Suppose one model executes $n$ rounds and records the average loss value of each round across different DL frameworks. Record two frameworks as $X$ and $Y$ and the two loss series are $X_{loss}=(x_1, x_2,..., x_n)$, and $Y_{loss}=(y_1,y_2,...,y_n)$. $D(i, j)$ is the DTW distance between the first $i$ elements of $X$ and the first $j$ elements of $Y$. The calculation of DTW is shown in the following formula:

\vspace{-4mm}
\begin{equation}
    D(i, j) = d(x_i, y_j) + min(D(i-1, j), D(i, j-1), D(i-1, j-1)) \\
    \label{equ:dtw} 
\end{equation}


In Formula (\ref{equ:dtw}), $d(x_i, y_j)$ represents the euclidean distance between $x_i$ and $y_j$ and 
We set $D(0, 0) = 0$, $D(i, 0)= \infty$ and $D(0, j) = \infty$ to ensure the correctness. We calculate $D (n, n)$ as the distance between $X_{loss}$ and $Y_{loss}$.

Besides, we record the time of the model across different frameworks during model training and inference. If the time of one framework far exceeds the other compared with the preset threshold, DevMuT reports an efficiency defect. 
We also compare the evaluation metric of the models across different frameworks and report a performance defect when the evaluation metric difference of models across different frameworks exceeds the preset threshold.

\noindent\textbf{Resource Defect.} The resources refer to the GPU and memory, and the defects are related to unreasonable scheduling, such as memory leaks. DevMuT records the size of memory and GPU used across different DL frameworks during execution and detects defects from the perspective of the three aspects:
(1) the used resources anomalously increase and lead to memory or GPU leak;
(2) the maximum used resource exceeds the preset threshold;
(3) the cosine similarity calculated by the memory or GPU size of the models across different DL frameworks is lower than the preset threshold. \looseness=-1

\noindent\textbf{Accuracy Defect.}
The accuracy is related to the model outputs in forward propagation and back-propagation. If the model outputs across different frameworks exceed the preset distance threshold, DevMuT reports an accuracy defect. We also adopt Chebyshev distance~\cite{cantrell2000modern} to measure the output distance, and its calculation is shown in Formula~\ref{equ:layerdis}.
Besides, if the model's output contains outliers on one framework but not on another or contains different outliers, DevMuT also indicates an accuracy defect.

\noindent\textbf{Crash Defect.} 
We detect the crashes exposed in the mutation process and model execution (i.e. training and inference). More specifically, if a model mutates successfully on one framework but fails on another or exposes crashes during the training or inference on one framework but not on another, we report a crash defect.

\section{Experiment design}
\label{sec:exp}

\subsection{Research Questions}
\label{sec:rqdesign}

Our study investigates the following three research questions:

\noindent\(\bullet\)
\textbf{RQ1: How does DevMuT perform compared to existing work?}
We evaluate the performance of DevMuT from the perspective of the diversity of the generated test inputs and detected defects compared to the existing methods.

\noindent\(\bullet\)
\textbf{RQ2: What kind of defects are detected by DevMuT?}
We analyze the defects detected by DevMuT and show the typical cases in detail, including the types, symptoms, and root causes.

\noindent\(\bullet\)
\textbf{RQ3: How does the search strategy contribute to DevMuT?}
We perform an ablation experiment to evaluate the double-Q search strategy's effectiveness and its contribution to DevMuT. \looseness=-1

\subsection{Benchmark}

The benchmark used in our study includes DL frameworks, DL models, and the corresponding datasets. The details are as follows.

\noindent\textbf{DL Frameworks.} We select the latest versions of three DL frameworks as the test objects, i.e., PyTorch 1.10.1, MindSpore 2.2.0, and JAX 0.4.27. 
PyTorch is a popular DL framework that supports flexible conversion to different mobile systems and dynamic computing graphs for convenience in development.
MindSpore is a new DL framework designed for various running environments with an efficient execution engine to enable users to manage computing resources and optimize task execution efficiently.
JAX provides powerful automatic differentiation and GPU/TPU support and is widely used in DL development and scientific computing.
Both three DL frameworks have been widely adopted in prior work~\cite{zou2023ramos,liu2023neuri,liu2023generation} and have active developer communities in the recent three years.

\noindent\textbf{DL Models and Datasets.} 
We collect 17 DL models from real industry scene tasks to ensure the practical meaning of our study. Specifically, VGG16, ResNet50, MobileNetv2, and VIT are four image classification models used in facial recognition.
The YoLoV3, YoLoV4, and two SSD models conduct object detection tasks and are widely applied in autonomous driving.
The Unet, Unet++, DeeplabV3, and DeeplabV3++ are semantic segmentation models for medical diagnosis.
The TextCNN conducts sentiment analysis on movie reviews. Lastly, SSIM-AE, PatchCore, OpenPose, and CRNN handle image-based defects, anomaly detection, key point detection, and scene recognition.
Besides, we also collect 12 DL models (labeled with ``*'' in Table~\ref{tab:expinfor}) from existing work~\cite{wang2020lemon,li2023comet,guo2020audee} to ensure the fairness of our study.
The details are shown in Table~\ref{tab:expinfor}. The first and second columns show the task and name of the DL models. The third to fifth columns show the parameter scale, model depth, and width. \looseness=-1

\begin{table}[]
  \centering
  \caption{Statistics of DL Models in Our Study}
   \vspace{-4mm}
  \resizebox{0.5\textwidth}{!} {
    \begin{tabular}{ccccc}
    \hline
    \textbf{Scene Task} & \textbf{Model} & \textbf{Parameters} & \textbf{Depth} & \textbf{Width} \\
    \hline
    \multirow{14}[2]{*}{Image classification} & VGG16-1 & 134,314,186 & 53    & 4,096 \\
          & ResNet50-1 & 23,581,642 & 126   & 2,048 \\
          & MobileNetV2-1 & 2,270,794 & 141   & 1,280 \\
          & VIT   & 87,423,754 & 150   & 3,072 \\
          & AlexNet* & 24,736,682 & 17    & 4,096 \\
          & LeNet5-1* & 107,764 & 10    & 120 \\
          & LeNet5-2* & 218,302 & 11    & 120 \\
          & ResNet50-2* & 123,914,152 & 126   & 2,048 \\
          & VGG19* & 143,661,736 & 45    & 4,096 \\
          & InceptionV3* & 1,254,998,216 & 26    & 1,000 \\
          & DenseNet121* & 7,045,504 & 248   & 1,000 \\
          & VGG16-2* & 138,353,320 & 39    & 4,096 \\
          & Xception* & 29,384,336 & 153    & 2,048 \\      
          & MobileNetV2-2* & 4,234,024 & 57    & 1,024 \\
    \hline
    \multirow{2}[1]{*}{Regression Prediction} & LSTM-1* & 81,301 & 7     & 1 \\    
          & LSTM-2* & 3,726 &  4     & 1 \\ \hline
    \multirow{4}[0]{*}{Object Detection} 
            & SSD-resnet50-fpn & 33,370,814 & 232   & 2,048 \\
          & SSD-mobilenetv1 & 11,329,461 & 143   & 1,024 \\
          & YoloV3 & 62,001,757 & 222   & 1,024 \\
          & YoloV4 & 65,741,916 & 334   & 1024 \\ \hline
    \multirow{4}[1]{*}{Semantic Segmentation} & DeeplabV3 & 58,149,077 & 263   & 2,048 \\
          & DeeplabV3++ & 59,453,824 & 272   & 2,048 \\
          & Unet  & 31,029,698 & 53    & 1,024 \\
          & Unet++ & 9,045,280 & 88    & 512 \\
    \hline
    Text Classification   & TextCNN & 859,146 & 12    & 96 \\
    \hline
    Anomaly Detection & PatchCore & 68,951,464 & 129   & 2,048 \\
    \hline
    Defect Detection & SSIM-AE & 2,720,768 & 39    & 500 \\
    \hline
    Key Point Detection & OpenPose & 52,311,446 & 112   & 512 \\ \hline 
   Scene recognition & CRNN  & 4,339,621 & 29    & 512 \\
    \hline
    \end{tabular}
    }
  \label{tab:expinfor}
       \vspace{-0.5cm}
\end{table}%

\subsection{Baseline Methods}

We select two state-of-the-art DL framework testing methods, i.e., COMET~\cite{li2023comet} and Muffin~\cite{gu2022muffin}. COMET is the latest mutation-based testing method with comprehensive types of mutation operators ranging from weight-mutation, input-mutation, and parameter-mutation to structure-mutation, which shares the same type as DevMuT. It aims to improve the middle layer coverage and generate new model structures with more diversity. Muffin generates new models based on the directed acyclic graph structure and detects defects in the model training stage, including forward calculation, loss calculation, and backward calculation, which shares the most common detection scope with DevMuT. 



\subsection{Measurements}
\label{sec:measures}

We evaluate DevMuT based on two measurements, i.e., generated mutants and detected defects.
For the first kind of measurement, we adopt three kinds of layer coverage to evaluate the diversity of the generated models adopted in previous work~\cite{li2023comet}:

\noindent\textbf{Layer Input Coverage (LIC).} Record the total number of data types as $N_{type}$, the total number of dimensions as $N_{dim}$, and the total number of shapes as $N_{shape}$. The number of data types covered is $n_{type}$, the number of dimensions covered is $n_{dim}$, and the number of shapes covered is $n_{shape}$. 
The layer input coverage $Input_{cov}$ is calculated as the result of the sum of $N_{type}$, $N_{dim}$, and $N_{shape}$, divided by the sum of $n_{type}$, $n_{dim}$, and $n_{shape}$. \looseness=-1

\noindent\textbf{Layer Parameter Coverage (LPC).} Record the number of different middle layers in the current generated models as $n_{ut}$, and the number of different edges in all the generated models as $N_{ut}$.
The layer parameter coverage is calculated as the result of $N_{ut}$ divided by $n_{ut}$. \looseness=-1

\noindent\textbf{Layer Sequence Coverage (LSC).} The layer sequence refers to the groups consisting of two different framework interfaces. Record the total number of all the interface groups that appeared in the execution as $AP_{sum}$ and the number of current generated models as $AP_{cur}$. The layer sequence coverage is calculated as the result of $AP_{sum}$ divided by $AP_{cur}$. \looseness=-1

Besides, we further adopt the Scott Knott ESD test~\cite{jelihovschi2014scottknott,chen2020different,chen2019software} to analyze the performance of $DevMuT$ and baseline methods on these three metrics. This statistical test method is often used to effectively identify significant differences and categorize the results of different methods into distinct classes.



For the second kind of measurement, we measure the defect detection ability of different testing methods based on the type and count of detected defects. Then we further count the number of reported defects, the confirmed defects, and the fixed defects in detail. \looseness=-1

\section{Result analysis}

\subsection{RQ1: Compared with Baseline Methods}
\label{sec:rq1result}


We evaluate DevMuT from two aspects: (1) the diversity of generated models and (2) the detected defects compared to baseline methods.
For the first aspect, we run DevMuT and baseline methods for 100 rounds followed by the experiment settings of existing work~\cite{wang2020lemon} and calculate the $LIC$, $LPC$, and $LSC$ metrics. Besides, we also count the illegal models that violate the mutation constraints.
For the second aspect, we analyze the defects detected by baselines exposed in our experiment and previous studies with those detected by DevMuT.
Besides, COMET cannot execute on DL models collected in our study (models are not labeled with ``*'' in Table~\ref{tab:expinfor}) for the following reasons.
(1) Its implementations cannot fit the new models since their structures are more complex than those previously used, thus always leading to crashes.
(2) Modifying the implementations to accommodate new models may lead to uncontrollable effects on COMET and compromise fairness.
Therefore, we only run DevMuT and COMET on those 12 DL models applied in existing work for fairness.

	
    


\begin{figure}[]
     \centering
    \includegraphics[width=0.95\linewidth]{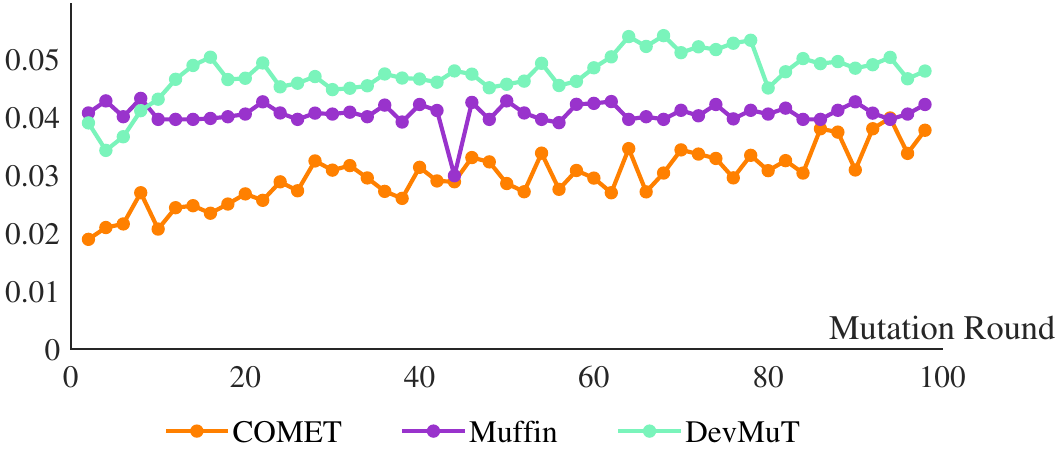}
    \vspace{-4mm}
     \caption{Comparison on the Diversity of Generated Models} 
          \vspace{-0.6cm}
     \label{fig:rq1_plot1}
\end{figure}

\begin{figure*}[]	
    \centering    
    \begin{subfigure}[b]{0.3\textwidth}
        \includegraphics[width=\textwidth]{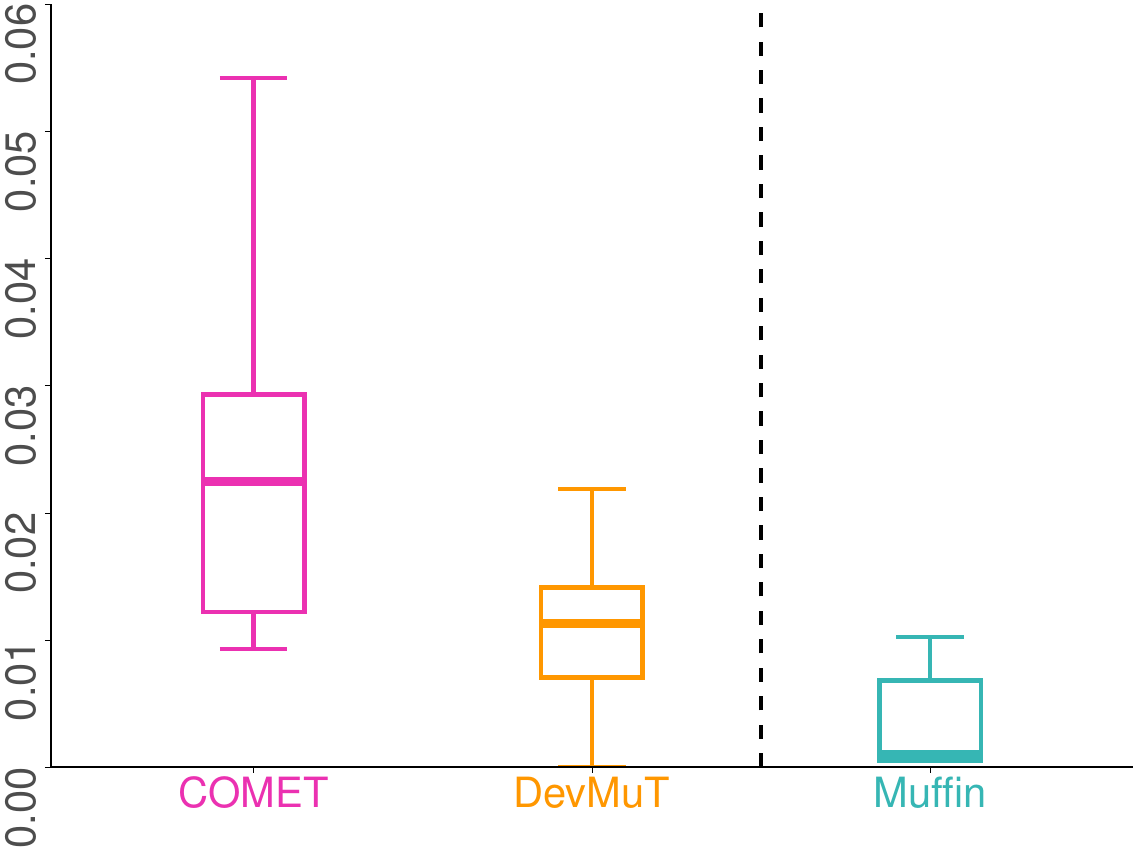}
        \vspace{-6mm}
        \caption{LIC}
        \label{fig:sklic}
    \end{subfigure}
    \begin{subfigure}[b]{0.3\textwidth}
        \includegraphics[width=\textwidth]{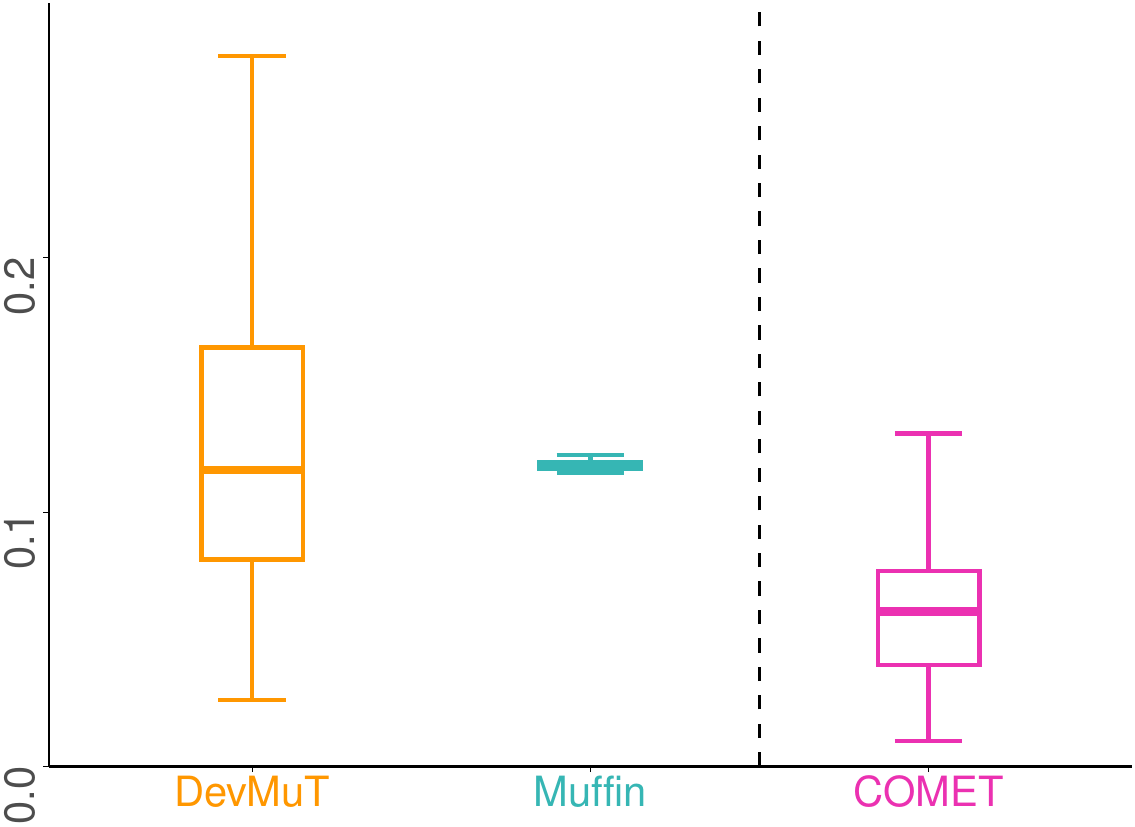}
        \vspace{-5mm}
        \caption{LSC}
        \label{fig:sklsc}
    \end{subfigure}
    \begin{subfigure}[b]{0.3\textwidth}
        \includegraphics[width=\textwidth]{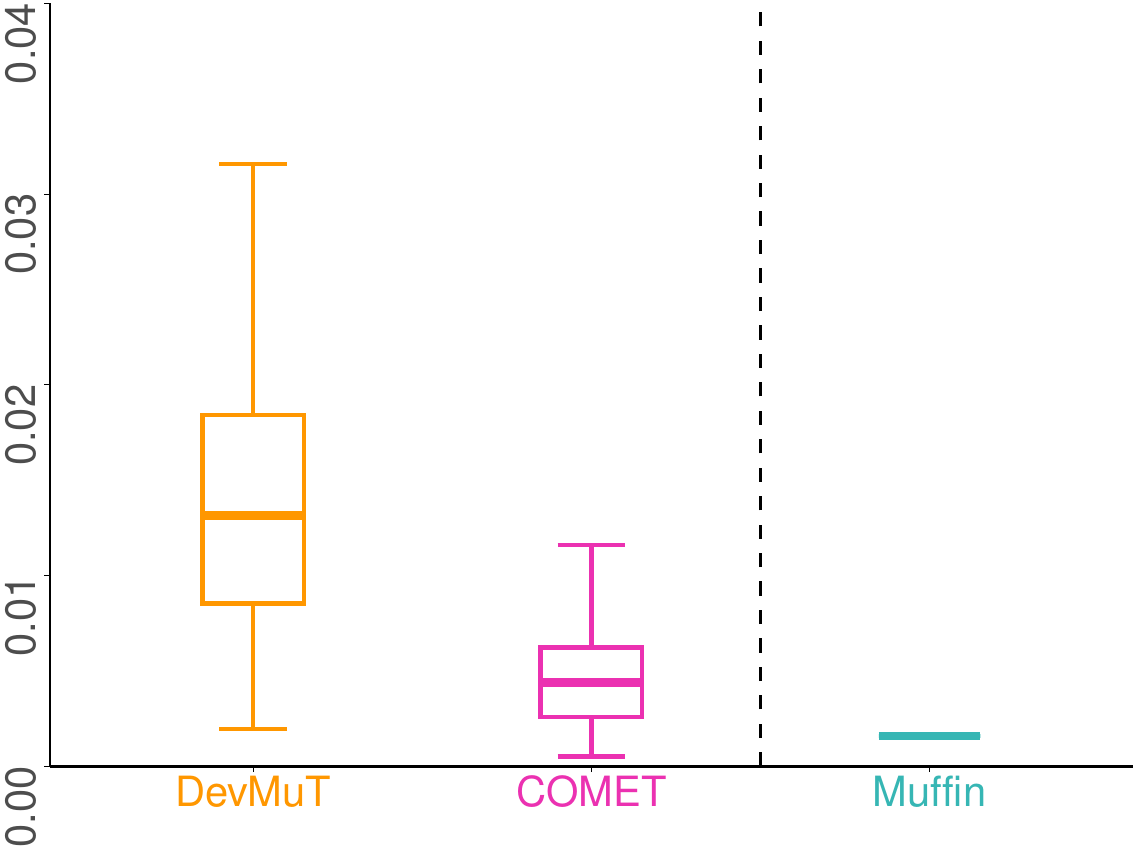}
        \vspace{-6mm}
        \caption{LPC}
        \label{fig:sklpc}
    \end{subfigure}
    \vspace{-5mm}
	\caption{Scott Knott Test Results on Three Diversity Metrics}
   \vspace{-5mm}
	\label{fig:sktest}

\end{figure*}

\noindent\textbf{Comparison based on the Generated Models.}
We show the average results in Figure~\ref{fig:rq1_plot1}. The $x$ axis represents the mutation rounds while the $y$ axis represents the average value of three diversity metrics of each generated model. 
Among the three diversity measurements, DevMuT outperforms the other two strategies in $LSC$ and ranks second in the other two metrics. However, DevMuT achieves the best performance based on the average of the three indicators as shown in Figure~\ref{fig:rq1_plot1}.
More specifically, DevMuT can introduce richer cascade operators (such as FPN structures) into the original model, thus increasing more diverse layer sequences.  
Since DevMuT follows the default setting of new structures, the parameter settings and input/output sequence are lower than Muffin and COMET, respectively. 
COMET combines the three metrics to select the mutation operator and seed model with no constraint, e.g., the value range of parameter mutation, so it can achieve higher $LPC$. 
However, the generated models of COMET greatly differ from the real-world model, with 36.69\% percent (434 illegal models in 1183 models) of models that violate the mutation constraints. DevMuT only generates 226 illegal models and achieves 81.17\% legal rates in total.
Muffin has a strong shape/dimension matching mechanism and can effectively avoid models with crashes. 
Meanwhile, it can generate more diverse groups of input/output sequences of middle layers, thus achieving higher $LIC$ than the other two methods. 
However, although the generated model does not violate the mutation constraints, they are only a combination of several DL operators and not actual models that can adapt to industrial tasks.
Besides, we conduct Scott Knott ESD test on these three metrics and show results in Figure~\ref{fig:sktest}. The $x$-axis represents different methods, while the $y$ -axis shows the values of the three metrics. 
The dashed line perpendicular to the $x$-axis separates methods with significant performance differences; methods on opposite sides show statistically significant differences, whereas those on the same side do not. In this figure, though DevMuT may not always achieve the highest model diversity, it maintains a comparable ability to generate diverse models while ensuring the model’s validity and practical significance, and its overall performance is superior to existing methods.\looseness=-1

\begin{figure}[]
  \centering
  \includegraphics[width=0.95\linewidth]{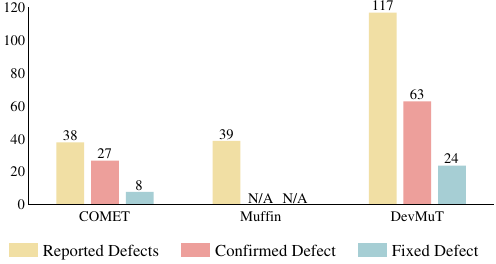}
        \vspace{-0.4cm}
  \caption{Number of Defects Detected by Different Methods}
  \vspace{-0.4cm}
  \label{fig:rq1defect}
    
\end{figure}

\noindent\textbf{Comparison based on the Detected Defects.} 
We count the defects detected by different methods and show the results in Figure~\ref{fig:rq1defect}. In particular, the $x$ axis represents the method name; the $y$ axis represents the counts of the total detected defects, the confirmed defects, and the fixed defects.
Specifically, COMET reports no inconsistency defects, four NAN defects, and six crash defects in our study after manual inspection. Four NAN defects and three crash defects receive confirmation. Besides, COMET previously reports 28 defects, of which developers confirm 20, including five inconsistency defects, three wrong output defects, one NAN defect, seven interface implementation defects, two conversion failure defects that are related to the conversion frameworks, and two ``Core Dump'' defects related to two DL operators.
Muffin does not run on new DL frameworks and reports 17 inconsistency detects, one NAN defect, and 21 crash defects in their previous studies. Since they do not report further confirmed information, we fix ``N/A'' on the count of confirmed and fixed defects of Muffin in Figure~\ref{fig:rq3_plot1}. Among all inconsistency defects, over 66.7\% are due to operator implementation errors, with only one defect related to the incorrect use of the ``BinaryCrossentropy'' loss function in TensorFlow. The back-propagation inconsistency defect stems from the improper implementation of the ``ReLU'' function. 
One NAN defect is attributed to an implementation error in the ``GlobalMaxPooling'' function while the remaining 21 crash defects have not been reported or analyzed in detail.
DevMuT detects 17 defects during execution, and 14 are confirmed. Except for one inconsistency defect and three NAN defects, DevMuT can detect three new types of defects (i.e., memory leak, efficiency and performance decrease during training, and crashes exposed in training) that cannot be detected by existing work. The number of three types is all two. The remaining six defects are related to the wrong implementations, unreasonable resource scheduling of DL operators, and other framework interfaces.\looseness=-1


\noindent\textbf{Answer to RQ1.} 
 (1) DevMuT achieves 71.68\% diversity average improvement on the generated models and achieves a higher 81.17\% legal rate exceeding baselines by 28.20\%.
 (2) DevMuT can detect three new kinds of defects that baselines cannot detect and shows a higher detection ability that meets the expectations of developers on defect detection.\looseness=-1

\subsection{RQ2: Analysis on Detected Defects}
\label{sec:rq2result}

\begin{figure}[]
     \center
      \includegraphics[width=1\linewidth]{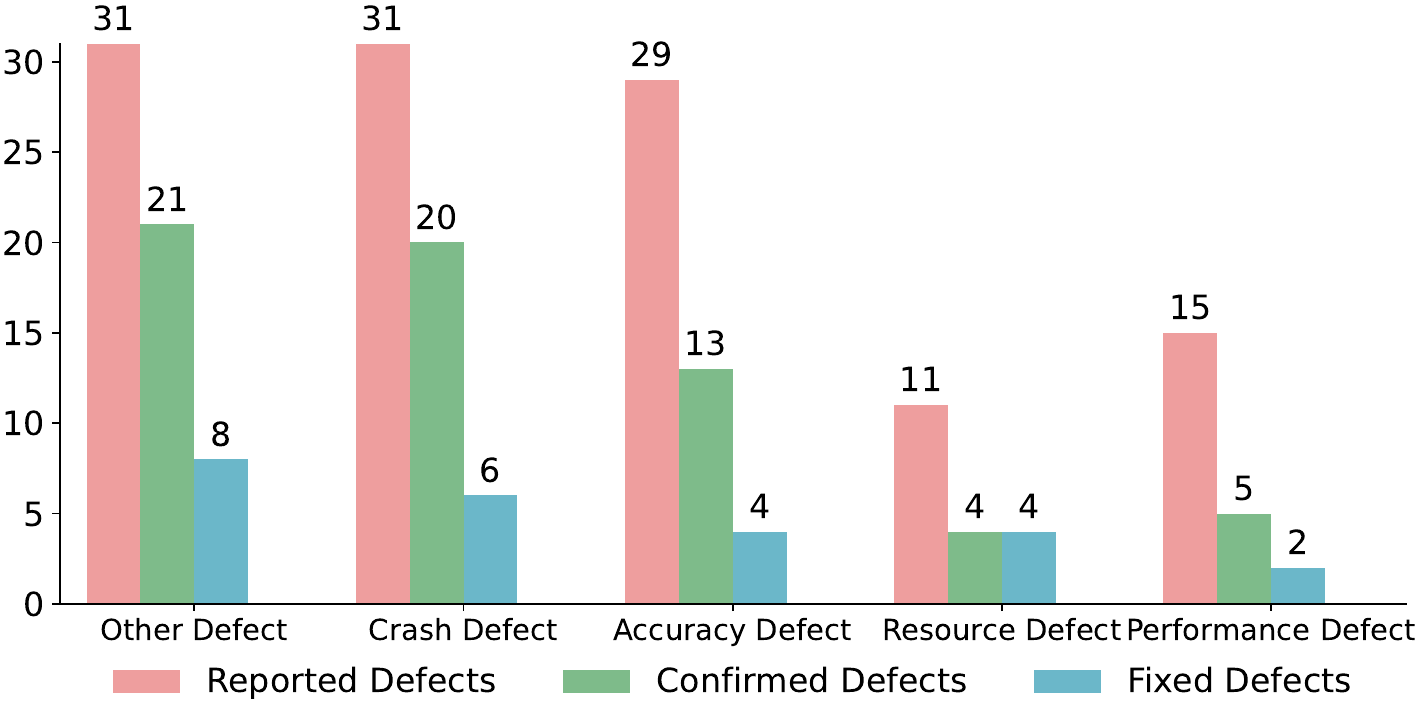}
      \vspace{-6mm}
     \caption{Defects Detected by DevMuT}
     \vspace{-6mm}
     \label{fig:rq2hpdefect}
\end{figure}

DevMuT detects 117 defects, and 63 are confirmed; eight are labeled with specific tags like ``Main'', and 24 defects have been fixed. Throughout the DL model lifecycle, which includes construction, execution, and deployment, we count seven defects (e.g., failure to initialize the middle layer) in the construction stage, 21 defects (e.g., abnormal loss) in the training stage, 12 defects (e.g., inference performance degradation) in the inference stage and 14 defects (e.g., crashes in pre-trained weight load) in the deployment stage among the 63 confirmed defects (the left 9 defects are document and framework functionality defects).
We use Figure~\ref{fig:rq2hpdefect} to show the defects detected by different test oracles: the $x$ axis represents the names of different types of defects; the $y$ axis represents the number of the total detected defects, the confirmed defects, and the fixed defects. The typical cases of each type are shown in the following part. \looseness=-1

\noindent\textbf{Performance Defect.} 
DevMuT detects 9 loss defects, 3 inference evaluation defects, and 3 efficiency defects. Firstly, we present a defect related~\cite{Performancedefect1} to NAN loss. Specifically, DevMuT mutates ``TextCNN'' for 20 rounds and detects the loss of MindSpore model changes from ``NAN'' to ``inf'' during the first train epoch while the loss of PyTorch model changes from 0.171 to 5.969. 
We find the output distances of the middle layers across different frameworks range from 1e-1 to 1e3 after training on 200 batches, indicating a defect in the framework interfaces related to weight updates. Developers have confirmed the defect report and replied to fix it. 
Next, we show two examples of the performance defect~\cite{Performancedefect2} in model inference and efficiency defect~\cite{Performancedefect3}, respectively.
The inference eval of the ``LeNet-5'' from MindSpore remains at 11.35\% with the loss hovering between 2.0-2.9 in the first 46 epochs. Then, it begins to increase to 30.60\% until it rises to around 99\% with the loss in 1e-5-1e-7. However, from the 164th epoch, although the accuracy was still around 98.64\%, the loss value changed to -inf, and then the loss value remained unchanged at NAN, with the accuracy plummeting to 9.8\%. After the analysis, developers find the implementation of the loss function, i.e., the ``CrossEntropyLoss'' contains defects. It is fixed in the latest version. 
Another example of an efficiency defect is ``Openpose''. The generated model from MindSpore is 45.86\% slower than that of PyTorch after 50 mutations. After preliminary analysis, we find the calculation time is mainly spent on the ``Flatten'' operator. Developers confirm that the ``Flatten'' operator of MindSpore lacks necessary optimization and is not supported in the current version. This defect will be fixed in future updates.

\noindent\textbf{Resource Defect.} Among the 11 defects, they can be subdivided into two types: (1) six are memory leaks caused by resource scheduling mechanisms during training, and (2) five are unreasonable memory allocations for single DL operators.
For the first type, we list an example~\cite{resourcedefect1} of the ``LeNet-5-1'' model to present. We find the memory usage rate slowly increases during the execution stage. Specifically, memory usage starts at 0.704GB, reaches 5.696GB after one day, and 9.408GB the next day. This is caused by continuous caching and failure to timely recycle unused memory by the MindSpore decorator ``ms.jit''.
Developers label it as a valuable defect report and decide to fix it with their highest priority. 
The next example~\cite{resourcedefect2} shows a typical case of the second defect type.
After 50 mutations of ``VGG16-1'', we find that the MindSpore model threw a memory allocation exception, while the equivalent PyTorch model normally executed because the ``Pad'' operator within the model cannot process the input data size. The developers confirmed that it is caused by the wrong implementations of the ``Pad'' operator on secure memory settings.

\noindent\textbf{Accuracy Defect.} 
DevMuT detects 29 accuracy defects, and 13 of them have been confirmed. Among the confirmed defects, nine are related to NAN outputs, and we list one typical example~\cite{accuracydefect1} related to the Ascend hardware. Specifically, the output of ``VGG16-1'' changes to NAN on MindSpore and PyTorch after six mutation rounds. However, when we reproduce it from GPU v100 to Ascend 910, the output changes from NAN to normal values. Developers detect the defect in the output design of Ascend 910 hardware. This defect is labeled with ``Main'' and achieves the focus of developers.\looseness=-1

\noindent\textbf{Crash Defect.} DevMuT detects 12 crash defects during mutation and 19 in execution. 
Defects in mutation arise from the implementation differences of various framework interfaces, leading to inconsistent results.
For example, when the kernel size exceeds the input data size, the ``Conv2d" operator in MindSpore can execute normally, whereas in PyTorch, it cannot~\cite{crashdefect1}.
Defects exposed in execution are often exposed in complex invoking scenarios.
One defect~\cite{crashdefect2} is exposed during the forward progress in model training.  For example, ``Vit'' failed to execute forward propagation after mutations. Developers locate the defect of the optimization mechanism in MindSpore. Since we select the lower optimization degree during execution, it cannot trigger ``constant folding'' and eliminate the redundant layer that leads to the ``slice error'' crash.

\noindent\textbf{Other Defects.} Except for the four kinds of defects detected by our designed test oracles, we detect 31 defects during our experiments. Specifically, we detect nine document defects about the wrong description of the framework interfaces and existing incomplete functionalities of frameworks.


\noindent\textbf{Answer to RQ2.}  DevMuT is good at detecting the crash defects exposed in complex invoking scenarios while also detecting new types of detects like resource scheduling defects and abnormal model performance defects, which shows better defect detection ability than existing methods.\looseness=-1

\vspace{-4mm}
\subsection{RQ3: Ablation Study on Mutation Guideline}
\label{sec:rq3result}
The mutation guideline is designed to determine the selection of seed models and mutation operators for DevMuT and can affect the performance. We conduct an ablation study to evaluate the contribution of the double-Q learning strategy to DevMuT. In particular, we implement two baselines named $DevMuT_{r}$ and $DevMuT_{mcmc}$ that select mutation operators and seed models by the random strategy and the MCMC strategy~\cite{brooks2011handbook}, respectively. These two strategies are widely applied in existing work~\cite{wang2020lemon,li2023comet} and show encouraging performance.
We run DevMuT, $DevMuT_{mcmc}$, and $DevMuT_{r}$ for 100 iterations and analyze the generated models from the perspective of model diversity and the number of legal ones. \looseness=-1

	
    


\begin{figure}[]
     \centering
    \includegraphics[width=0.95\linewidth]{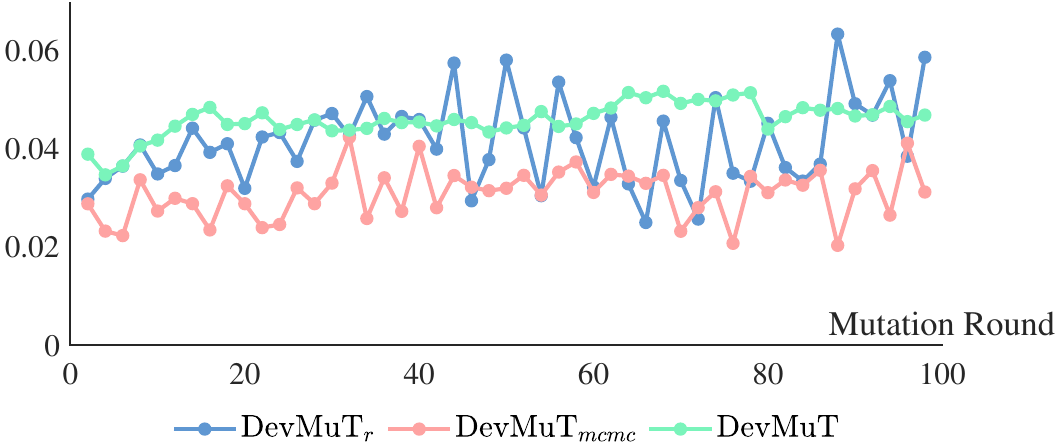}
     \vspace{-0.3cm}
     \caption{Model Diversity Comparison on Different Strategies}  
     \vspace{-0.5cm}
     \label{fig:rq3_plot1}
\end{figure}

We show the average results for all models in Figure~\ref{fig:rq3_plot1}. The $x$ axis represents the mutation rounds while the $y$ axis represents the average value of three diversity metrics of each generated model.
As shown in Figure~\ref{fig:rq3_plot1}, 
$DevMuT_{r}$ can reach the peak earlier than the other two strategies, but as the mutation deepens (after 20 rounds of mutation), it is prone to falling into ``local optima'' and often produces illegal models, and the performance is no longer stable; 
The mutation process of $DevMuT_{mcmc}$ is relatively stable, with the variance on three metrics being the smallest among the three strategies. The generated model with maximum value on three metrics is, on average 15.04\% and 14.86\% lower than $DevMuT_{r}$ and $DevMuT_{mcmc}$, respectively. The model generated by DevMuT gradually increases with execution, reaching a peak between 60-80 rounds before slowly decreasing. Compared with the other two strategies, the model generated by DevMuT has better average values of $LSC$, but its stability is slightly lower than $DevMuT_{mcmc}$.
Besides, DevMuT achieves the highest legal rate, 81.17\%, and outperforms $DevMutT_{mcmc}$ and $DevMutT_{r}$ with 73.83\% and 51.92\% among all the generated models across three strategies, respectively.
The average time of DevMuT, $DevMuT_{mcmc}$, and $DevMuT_{r}$ to generate each model is 95.08s, 97.92s, and 244.917s since $DevMuT_{r}$ often generate illegal models and need to spend extra time on backtrack to explore previously test input space. \looseness=-1

\noindent\textbf{Answer to RQ3.} Results show that double-Q learning performs better than the random and MCMC strategy regarding the diversity and legitimacy of generated models and execution efficiency.

\vspace{-2mm}
\section{Discussion}
\label{sec:discussion}

This section discusses DevMuT's performance in efficiency and false positives. Its computational cost includes (1) model mutation and (2) defect detection. For model mutation, the average times for DevMuT, COMET, and Muffin are 95.08, 98.18, and 14.45 seconds, respectively. DevMuT takes longer than Muffin due to its enhanced validation process, which improves legal rates of generated models by 28.20\%, improving test effectiveness. 
For defect detection, DevMuT requires around 45 minutes per model, compared to a few seconds for baselines, but it identifies four new defect types: abnormal loss, low training efficiency, training crashes, and unreasonable resource scheduling that are valuable to developers with eight confirmed as ``Serious'' by developers. Baseline methods, which focus on accuracy defects during inference, fail to detect these defects.
We argue that the execution time of DevMuT is not a concern in practice, especially in industry: 
(1) the critical nature of detected defects justifies the testing, as undetected defects may cause longer delays; 
(2) the process is fully automated and can run in backgrounds; 
(3) advancements in AI accelerators (e.g. GPU~\cite{gpu}, NPU~\cite{npu}, and TPU~\cite{tpu}) can rapidly improve efficiency.

To deal with false positives, DevMuT employs nine mutation constraints to avoid and filter illegal models, supplemented by manual inspections for the remaining cases. The first six constraints are designed to guide mutation with the development experience of developers. The second three leverage tactics in identifying and filtering illegal models, reducing the false positive rate to 22.22\%.
Remaining false positives often include accuracy defects like loss inconsistency or crashes due to incorrect parameters, migration failures, or unreasonable configurations. For loss defects, if DevMuT reports loss inconsistency that does not affect inference, developers reject it. Crashes caused by missing settings are addressed by experienced developers, while other false positives are filtered via manual inspection. Experiments show these measures effectively handle false positives in DevMuT's defect reports.\looseness=-1

\vspace{-2mm}
\section{Threats to Validity}

\textbf{Internal Validity.} It mainly comes from the correctness of implementation. To address this, three authors alternately conduct cross-validation to ensure the implementation is correct and compare the original DL models' equivalence across different frameworks.\looseness=-1

\noindent\textbf{External Validity.} It mainly comes from the benchmark: DL frameworks and models may affect the generality of our study. To address this, we plan to encompass other frameworks like Jittor~\cite{jittor} and TensorFlow~\cite{tensorflow} to enhance our study's breadth. Besides, future research will expand to new model tasks, such as audio recognition. \looseness=-1

\noindent\textbf{Construct Validity.}  It mainly comes from the experiment parameter and interview guideline settings in the experiments and may affect the reliability of experimental results. For the parameter settings, we follow the recommendation of previous studies~\cite{gu2022muffin,li2023comet} when running baseline methods. Besides, we set the threshold of the test oracles and the parameters of DevMuT based on the experimental results to reduce the false positives and achieve better performance. 
For the interview, we send the collected interview results to the interviewees for confirmation. \looseness=-1

\noindent\textbf{Conclusion Validity.} It mainly comes from the rationality of evaluation metrics and may affect the effectiveness of our conclusions in real scenes. To address this, we follow previous studies' metrics for evaluation and conduct manual inspections to exclude false positives and enhance robustness.\looseness=-1

\section{Conclusion}
In this study, we propose a mutation-based DL framework
testing method DevMuT based on developers' expertise. 
To detect defects that are more relevant to real scenes, DevMuT designs seven mutation operators and nine extra constraints to simulate the common operations of developers on DL models and expand the detection scope to more stages of the model lifecycle (i.e., model training and inference).
DevMuT further utilizes the double-Q learning strategy to guide the mutation. 
Our comprehensive evaluation shows the effectiveness and efficiency of our proposed method.

\section*{acknowlegement}
We would like to thank the anonymous reviewers for their insightful comments. This work is supported partially by the National Natural Science Foundation of China (61932012, 62372228, 62141215), Science, Technology, and Innovation Commission of Shenzhen Municipality (CJGJZD20200617103001003).

\bibliographystyle{ACM-Reference-Format}
\bibliography{sample-base}

\end{document}